Beitragsnummer:1185

# Inverted Classroom in der Einführungsveranstaltung Programmierung

Ulrich von Zadow, Natalie Kiesler

Technische Hochschule Nürnberg Georg Simon Ohm

## Zusammenfassung

Die Lehrveranstaltung Programmierung für Informatik-Studierende wurde an der TH Nürnberg, wie an vielen Hochschulen, bisher als klassische Vorlesungs-Übungs-Kombination realisiert. Nach einer hohen Durchfallquote im Wintersemester 2023/24 wurde im Wintersemester 2024/25 für eine Kohorte ein experimentelles, auf dem Inverted Classroom-Ansatz basierendes Lehrkonzept implementiert. Hierbei bereiten sich die Studierenden durch Literaturarbeit auf die Veranstaltungen vor, in denen vorwiegend aktivierende Lehr-/Lernformen genutzt werden. Die Veranstaltung wurde durch eine Reihe von Datenerhebungen (Teaching Analysis Poll, zwei Umfragen, Lehrtagebuch) begleitet, um das Lernverhalten der Studierenden, beobachtbare Hürden, und Good Practices für zukünftige Veranstaltungen abzuleiten. Das Konzept wurde insgesamt positiv bewertet, wobei sich auch viele Verbesserungsmöglichkeiten im Detail abzeichneten. Im Artikel dokumentieren wir die Ergebnisse der Erhebungen und diskutieren die Implikationen.

## 1. Einleitung

Das Erlernen von Programmieren ist kognitiv anspruchsvoll und stellt eine zentrale Hürde in Informatikstudiengängen dar (Du Boulay, 1986; Gomes und Mendes, 2007; Kiesler, 2022; Luxton-Reilly, 2016; Luxton-Reilly et al., 2018; Whalley, 2007; Xingalos, 2016). Dies spiegelt sich u.a. seit Jahren in hohen Abbruch- und Durchfallquoten wider (Heublein und Wolter, 2011; Heublein, 2017; Heublein et al., 2018; Heublein et al., 2020; Neugebauer et al., 2019). Die TH Nürnberg ist hier keine Ausnahme. Die Veranstaltung 'Prozedurales Programmieren' wird im ersten Semester der Informatik-Studiengänge seit längerem als klassische Vorlesungs-Übungs-Kombination gelehrt. Programmieraufgaben, die im Laufe des Semesters als Hausaufgabe gelöst werden, bilden die Zulassungsvoraussetzung zur Klausur, in der Programmieraufgaben auf Papier bearbeitet werden.

Im Wintersemester 23/24 hatte Generative AI hinreichend Bekanntheit erlangt, sodass Hausaufgaben damit gelöst werden konnten. Dies korrelierte mit sehr hohen Durchfallquoten (Kiesler et al., 2024) und führte dazu, dass im darauf folgenden Wintersemester mit der Kohorte von 67 Studierenden der Medieninformatik (Dozent Prof. Dr. von Zadow, zwei Übungsgruppen) ein experimentelles, auf dem Inverted Classroom-Ansatz basierendes, Lernkonzept gewählt wurde. Die klassische Vorlesung entfällt; Studierende bereiten sich durch Literaturarbeit selbstständig auf die Übungseinheiten vor. Diese werden für aktivierende Lehr-/Lernformen (u.a. Quizfragen und Programmieraufgaben) genutzt.

Wir sind der Frage, wie sich das neue Konzept in der Praxis bewährt, mit Hilfe von zwei Umfragen und einer Teaching Analysis Poll (Franz-Özdemir et al., 2019) nachgegangen. Zusätzlich wurde ein Kurstagebuch geführt. Konkret wurde die Untersuchung durch folgende Leitfragen bestimmt:

1. **Lernverhalten**: Welches Lernverhalten zeigen die Studierende? Wie subjektiv erfolgreich sind sie damit?

2. **Hürden**: Wie wirkt sich das Konzept auf klassische Hürden der Programmierausbildung (z.B. Debugging, IDE benutzen) aus?
3. **Lessons Learned**: Welche Good Practices ergeben sich? Wie könnte das Konzept weiter verbessert werden?

In diesem Artikel beschreiben wir zunächst das eingesetzte Lehr-Lernkonzept (Abschnitt 2). In den Abschnitten 3 und 4 widmen wir uns der Forschungsmethodik sowie den Ergebnissen. Abschließend diskutieren wir die Resultate und setzen sie in Relation zu den Forschungsfragen.

## 2. Lehr-/Lernkonzept

Der Inverted Classroom hat in den vergangenen Jahren an Popularität gewonnen (Strelan et al., 2020). Dieses Lehrkonzept ist dadurch gekennzeichnet, dass die Phase der reinen Wissensvermittlung (an Hochschulen üblicherweise als Vorlesung realisiert) außerhalb der gemeinsamen Präsenzzeit – beispielsweise über zur Verfügung gestellte Videos – stattfindet (Sams, 2012). Die synchrone Unterrichtszeit wird genutzt, um das Wissen anzuwenden und zu vertiefen. Diese Aufteilung schafft deutlich mehr Zeit für aktiven Unterricht und individuelle Förderung; die Lehrperson übernimmt die Rolle eines Coaches, anstatt Inhalte zu präsentieren (Lage et al., 2000). Eine Herausforderung des Modells ist der hohe Vorbereitungsaufwand für die Lehrkraft. Zudem ist der Erfolg davon abhängig, dass die Lernenden sich tatsächlich vorbereiten.

Im vorliegenden Kurs wurde das Semester zweigeteilt. In der ersten neunwöchigen Phase wurden die grundlegenden Konzepte des Moduls vermittelt und geübt. Die verbleibende Zeit wurde für vertiefende praktische Übungen genutzt. Das Lehrkonzept für die erste Phase (Semesterwochen 1-9) sah vor, dass sich die Studierenden durch Literaturarbeit auf die Übung (zwei Doppelstunden pro Woche) vorbereiten. Neben eigenen Texten wurde hierbei ein Buch zu C# als Quelle genutzt. In der ersten Doppelstunde wurde i.d.R. ein Multiple-Choice-Quiz zu den erlernten Konzepten durchgeführt, während in der zweiten Doppelstunde Programmieraufgaben gestellt wurden. Wir zielten hierbei auf eine natürliche Progression: Die Literaturarbeit legt das Fundament, die Quizze sichern das grundlegende Verständnis der Konzepte, und die Programmieraufgaben dienen der Vertiefung durch Anwendung und Transferleistungen.

Wichtig ist hierbei, dass das Quiz nicht als Prüfung konzipiert ist, sondern auf Verständnis abzielt. Er wird dementsprechend nicht benotet und im Think-Pair-Share-Format (Spannagel und Spannagel 2013) durchgeführt. Hierbei lösen die Studierenden im ersten Schritt alle Aufgaben allein (*Think*), im zweiten Schritt einigen sie sich in Paaren auf Antworten (*Pair*). Die Lehrkraft unterstützt in beiden Phasen, wenn sich Lücken zeigen. Fragen, die für viele schwierig waren, werden abschließend an der Tafel erklärt (*Share*). Das Verständnis wird hierbei also zunächst durch die eigenständige Beschäftigung mit dem Thema, anschließend durch Diskussionen und gegenseitiges Erklären, und bei Bedarf durch Unterstützung der Lehrperson gefördert.

Die Programmieraufgaben in der zweiten Doppelstunde dienen der praktischen Anwendung und Vertiefung der Konzepte. Im Gegensatz zu den Quizzen werden die Aufgaben i.A. am Ende der Übung abgegeben und anschließend benotet. Das Erreichen einer Mindestpunktzahl ist Zulassungsvoraussetzung für die Klausur. Zusätzlich wurde in der 6. Woche ein einstündiger Workshop zu Lernstrategien abgehalten, in denen die Studierenden in Kleingruppen ihre bisherigen Erfahrungen reflektieren sollten.

In der zweiten Phase des Semesters (Semesterwochen 10-14) wurden etwas größere Programmieraufgaben gestellt, deren Abgabe nicht verpflichtend war. Ziele waren die Vertiefung grundlegender Programmierkonzepte, das Schließen von Wissenslücken, sowie Transferleistungen. Da sich zu diesem Zeitpunkt schon die sehr unterschiedlichen Wissensstände der Studierenden gezeigt hatten, wurde eine relativ große Auswahl an Aufgaben mit verschiedenen Schwierigkeitsgraden zur Verfügung gestellt.

Das Ziel des vorgestellten Lehrkonzeptes war es, die bekannten Vorteile der Inverted Classroom-Methodik auf unsere Veranstaltung zu übertragen. Dies erschien uns insbesondere deshalb erfolgversprechend, da es sich in wesentlichen Anteilen um das Erlernen von prozeduralem Wissen handelt und daher ein Fokus auf praktische Übungen und individuelles Feedback sinnvoll erscheinen (Scott, 2003).

## 3. Methoden

Daten zu der Veranstaltung wurden durch ein Teaching Analysis Poll und zwei Umfragen sowie durch ein Kurstagebuch erhoben.

### 3.1 Teaching Analysis Poll

In Semesterwoche 9 wurde mit einer der zwei Übungsgruppen (18 Teilnehmende) ein Teaching Analysis Poll (TAP, Franz-Özdemir et al., 2019) durchgeführt. Moderiert von einer neutralen Person (in unserem Fall eine Mitarbeiterin der hochschulinternen Einrichtung für Lehr- und Kompetenzentwicklung) und in Abwesenheit der Lehrperson wurden in Kleingruppen unterstützende und hinderliche Faktoren für den Lernprozess sowie Verbesserungsvorschläge diskutiert. Die Ergebnisse wurden anonymisiert an die Lehrperson zurückgemeldet.

### 3.2 Umfragen

Im Verlauf des Semesters wurden zwei Umfragen zur Evaluation des Kurses durchgeführt. Beide Umfragen erfolgten anonym und unter Einholung eines Informed Consent der Teilnehmenden. Die Datenerhebungen fanden in den 10. und 13. Wochen des Semesters jeweils während der regulären Präsenzveranstaltungen statt. Beide Umfragen enthielten Fragen zur Selbsteinschätzung des Wissensstands sowie zu den praktischen Fertigkeiten der Studierenden. Die Antworten wurden auf einer vierstufigen Likert-Skala erfasst. Zusätzlich bestand für die Teilnehmenden die Möglichkeit, offene Kommentare abzugeben. Die zweite Umfrage umfasste zusätzlich Fragen zur Nutzung von Lernressourcen, zur Einschätzungen der Studierenden zu den Quizzen und Praktikumsaufgaben, sowie zum vorhandenen Vorwissen der Studierenden.

### 3.3 Kurstagebuch

Im Rahmen des Kurses wurde von der Lehrperson ein Kurstagebuch geführt, in dem nach jeder Übung die Eindrücke dokumentiert wurden. Dazu gehörten auch umfassende Aufzeichnungen während der Übung zu Lernstrategien in Semesterwoche 6.

## 4. Ergebnisse

Im Folgenden dokumentieren wir die Ergebnisse der Datenerhebungen.

### 4.1 Teaching Analysis Poll

Die Ergebnisse des Teaching Analysis Polls erwiesen sich aus formativer Sicht als sehr wertvoll. Teile der Befunde sind allerdings für die Einschätzung der Veranstaltung nur bedingt von Interesse, sodass wir hier eine verkürzte Darstellung wählen. Insbesondere der Einsatz von Quizzen als Lernmethode, die Möglichkeiten zur Zusammenarbeit mit Kommilitonen, sowie die individuelle Unterstützung durch die Lehrkraft wurden positiv bewertet. Gleichzeitig wurde das eingesetzte Buch deutlich kritisiert. Zudem wurde berichtet, dass für die erste Iteration des Quizzes oft nicht genug Zeit zur Verfügung stand.

## 4.2 Umfrage 1

Die Ergebnisse der ersten Umfrage, an der 32 Studierende teilnahmen, sind in Abb. 1 dargestellt. Da die Umfragen direkt in den Übungen durchgeführt wurden, ist mit Verzerrungen (Bias) zu rechnen. Die gewonnenen Daten sollten daher nicht überbewertet werden. Aus diesem Grund wurde auch auf eine statistische Auswertung verzichtet.

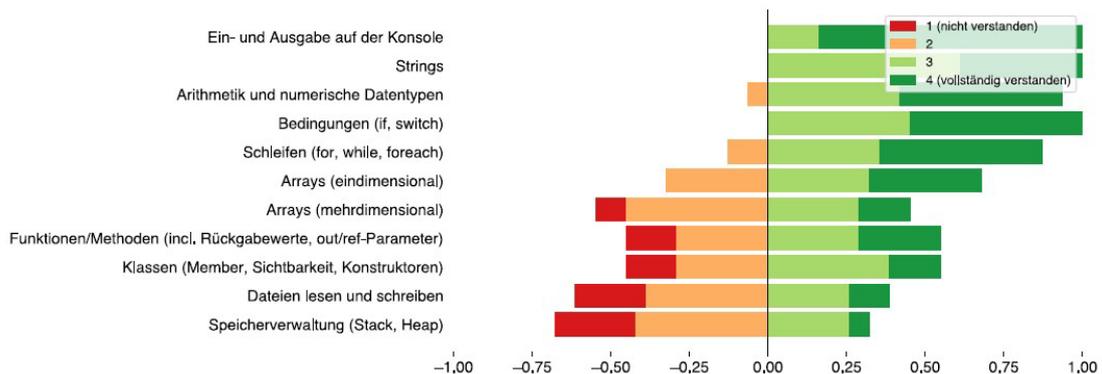

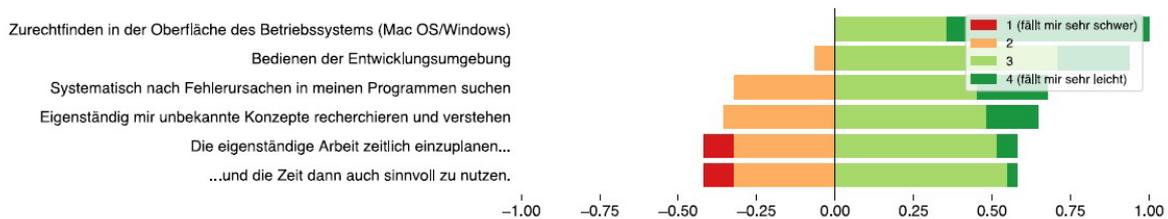

*Abb. 1: Ergebnisse Umfrage 1 (N=32)*

Hinsichtlich des Wissensstands zeigte sich, dass die meisten Teilnehmer die grundlegenden Konzepte bis einschließlich der Schleifen zu diesem Zeitpunkt verstanden hatten. Gleichzeitig wurde eine sehr große Bandbreite im Verständnis sowie bezüglich der Fertigkeiten im Bereich der Lerntechniken deutlich. Ein erheblicher Anteil der Studierenden gab an, Schwierigkeiten bei der eigenständigen Recherche sowie in der Zeitplanung und -nutzung zu haben.

## 4.3 Umfrage 2

An der zweiten Umfrage nahmen 20 Studierende teil. Auch hier gilt: Aufgrund der Durchführung in der Übung ist mit einem Bias in den Antworten zu rechnen. Hinsichtlich des Wissensstands und der praktischen Fertigkeiten ergaben sich ähnliche (wenn auch etwas verbesserte) Befunde verglichen mit der ersten Umfrage (Abb.2). Die Angaben zur Qualität der von den Studierenden geleisteten Vorbereitungsarbeiten sind insgesamt leicht positiv. Als Quelle wird das Buch wieder negativ bewertet, während andere Wissensquellen (durch den Dozenten geschriebene Dokumente, ChatGPT, weitere Netzressourcen) deutlich besser abschneiden (Abb. 3).

Betrachten Sie die folgenden Themen und schätzen Sie ab, wie gut Sie die Konzepte verstanden haben.
Beantworten Sie auf einer Skala von 1 bis 4 (1=Kein Verständnis, 4=Sehr gutes Verständnis).

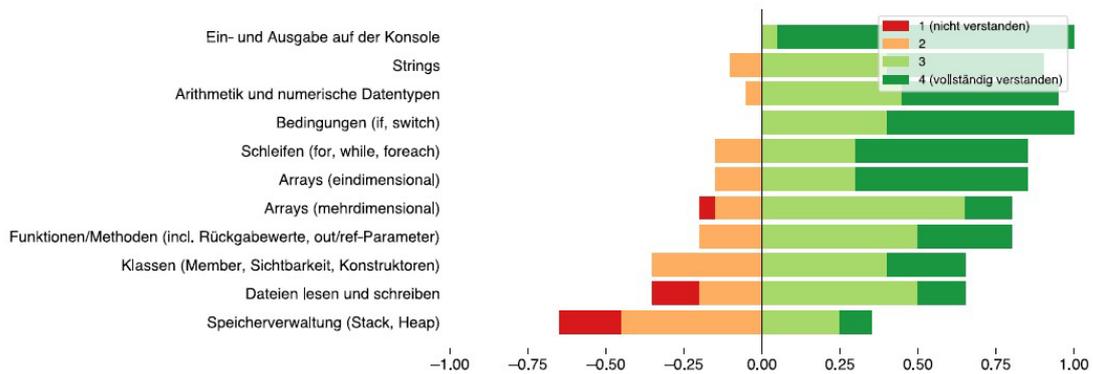

Schätzen Sie Ihre praktischen Fertigkeiten ein und geben Sie auf einer Skala von 1 bis 4 an, wie schwer bzw. leicht Ihnen folgende Tätigkeiten fallen:

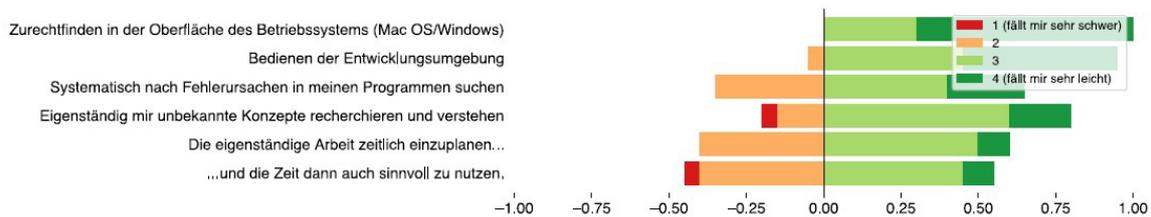

*Abb. 2: Ergebnisse Umfrage 2 (N=20) – Wissensstand und praktische Fertigkeiten*

Betrachten Sie die Einarbeitung in Themen der Programmierung, die Sie als Hausaufgaben bekommen haben.

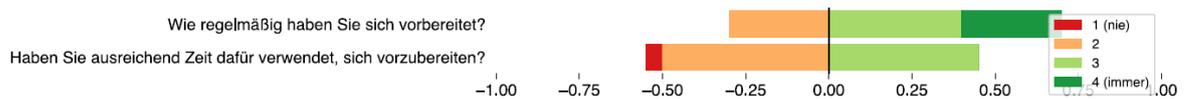

Wie oft haben Sie folgende Quellen genutzt?

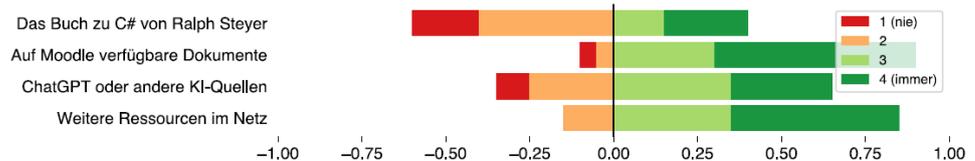

Wie hilfreich waren folgende Quellen?

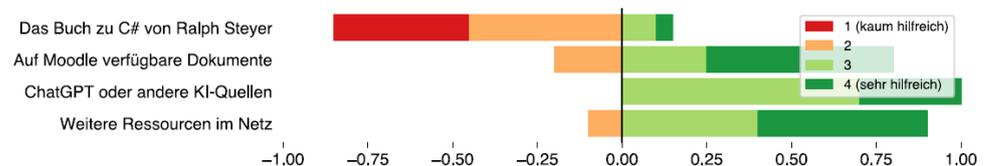

*Abb. 3: Ergebnisse Umfrage 2 (N=20) – Hausaufgaben und Quellen*

Insbesondere ChatGPT wird häufig genutzt und als sehr hilfreich eingeschätzt. Die Quizze wurden durchgängig positiv bewertet. Bemerkenswert dabei ist, dass die Einschätzungen zur Bearbeitungszeit eine große Bandbreite umfassen und – im Gegensatz zu den Ergebnissen des Teaching Analysis Polls – nicht einheitlich als zu knapp empfunden wurden. Der Schwierigkeitsgrad der Programmieraufgaben wird als angemessen beurteilt, wobei einige die Bearbeitungszeit als deutlich zu kurz einschätzten (Abb. 4).

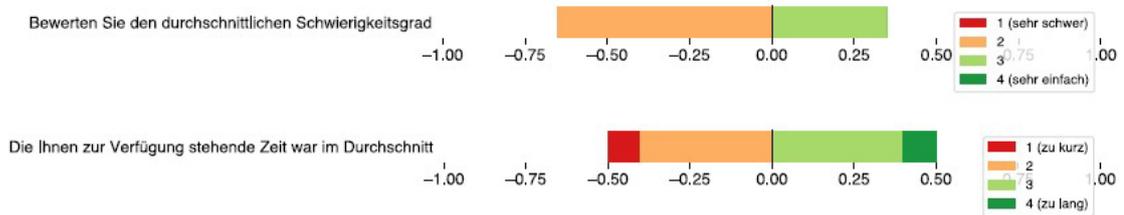

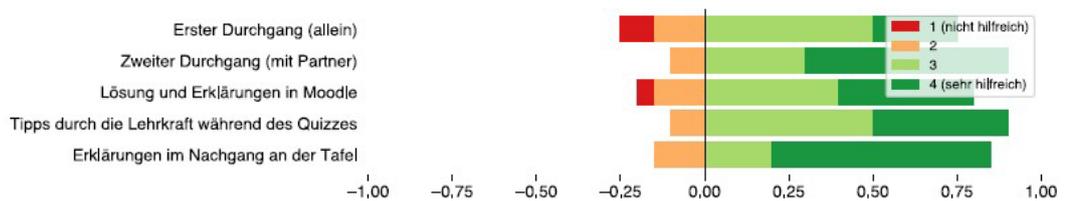

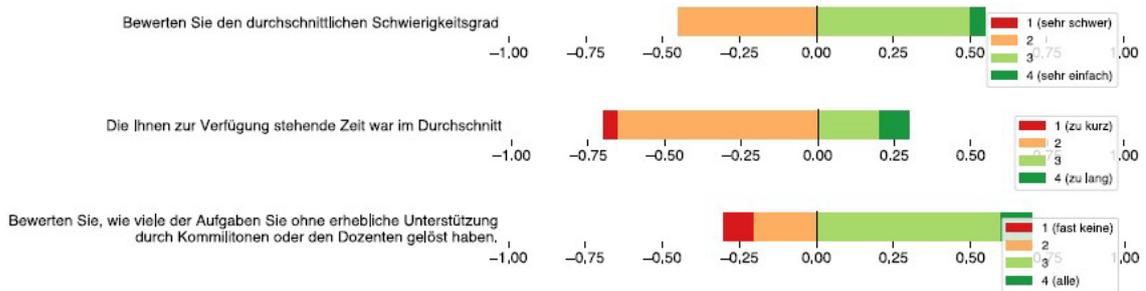

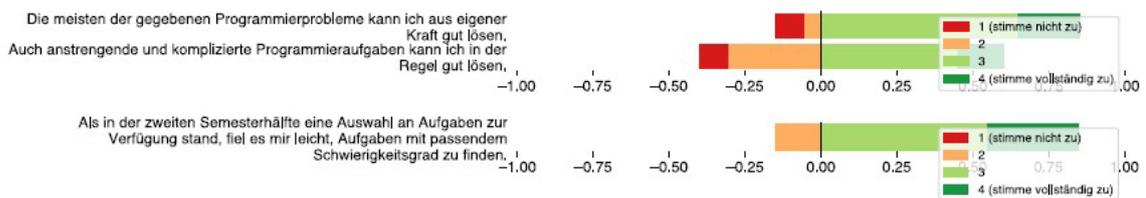

*Abb. 4: Ergebnisse Umfrage 2 (N=20) – Aktivitäten im Kurs*

## 4.4 Kurstagebuch

Im Kurstagebuch wurden zahlreiche Erkenntnisse zu Methoden und Aufgaben dokumentiert, die eher formativ von Interesse sind. Für die Forschungsfragen interessant sind im einzelnen folgende Ergebnisse. Beobachtungen zum Lernfortschritt ergaben, dass einige Studierende schon zu Beginn des Semesters mit dem Tempo überfordert waren. Gegen Mitte Dezember zeigte sich dies in einer großen Bandbreite an Kompetenzen: Während einige noch Schwierigkeiten bei den einfachsten Aufgaben hatten, konnten andere bereits komplexere Projekte erfolgreich bearbeiten. Mehrfach wurde beobachtet, dass Studierende zwar die Grundkonzepte verstanden hatten, aber Schwierigkeiten dabei hatten, diese zu Lösungen zu synthetisieren (z.B.: Ein Studierender erkennt korrekt, dass für eine Lösung sowohl eine if-Bedingung als auch eine for-Schleife nötig sind, setzt die Bedingung aber außerhalb statt innerhalb der Schleife).

Der Workshop zu Lernstrategien brachte weitere Erkenntnisse. Z.B. wurde das Buch ein weiteres Mal negativ bewertet. ChatGPT wird von nahezu allen Studierenden als Lernressource herangezogen. Dabei nannten viele das Stellen von Verständnisfragen sowie das Erklären von Codezeilen als wirksame Strategien, während das Generieren von Fragen und das Überprüfen von Antworten nur vereinzelt (dann allerdings positiv) erwähnt wurden.

## 5. Diskussion

Insgesamt wird das Konzept positiv bewertet, wenngleich auf Detailebene noch Verbesserungspotenziale bestehen (Kommentar in Umfrage 1: ‚Kurs hat in der Struktur viel Potential, muss aber noch verbessert werden.'). Offensichtlich scheint vor allem, dass das Buch vollständig durch eigene, auf das Curriculum abgestimmte, Texte ersetzt werden sollte. Im Folgenden werden die Leitfragen anhand der erhobenen Ergebnisse diskutiert.

### 5.1 Lernverhalten

Sowohl das Kurstagebuch als auch die Umfrageergebnisse deuten darauf hin, dass die Studierenden in der Regel relativ gut vorbereitet waren. Gleichzeitig sind die nicht sehr guten Ergebnisse der Umfragen im Bereich eigenständige Recherche sowie Zeitplanung und -nutzung ein Hinweis darauf, dass eine gezielte Förderung dieser Kompetenzen im Kurs hilfreich sein dürfte. Vor dem Hintergrund bestehender Vorarbeiten überrascht dies wenig (Kiesler, 2022).

Ebenfalls wenig überraschend scheint sich Generative AI als Assistenztool durchzusetzen. ChatGPT wird laut Umfrage von einer überwältigenden Mehrheit der Studierenden genutzt. Gleichzeitig scheinen Studierende hier bereits Kompetenzen aufzubauen: In den Veranstaltungen wurde recht differenziert über die Möglichkeiten und Grenzen diskutiert. Dies steht in Einklang mit anderen Erhebungen (Scholl und Kieser, 2024; Prather et al., 2023).

### 5.2 Hürden

Die wichtigste Erkenntnis im Bereich der Hürden ist, dass sich bereits früh im Kurs eine deutliche Kluft zwischen den leistungsstärksten und weniger erfolgreichen Studierenden abzeichnet. Ein zentraler Schwierigkeitsfaktor könnte darin liegen, dass Lernende gleichzeitig sowohl die Syntax als auch die Semantik von Programmiersprachen erfassen müssen – gerade am Anfang des Studiums eine erhebliche Herausforderung (Du Boulay, 1986). Visual Programming Languages wie Scratch können hierfür einen Lösungsansatz bieten, da sie die syntaktische Komplexität zunächst eliminieren (Bentrad und Meslati, 2011; Kiesler, 2016; Malan und Leitner, 2007; Zhang et al., 2013). In jedem Fall scheint es sinnvoll, in den ersten Wochen den kognitiven Anspruch zu reduzieren.

### 5.3 Good Practices

Aus den Ergebnissen lassen sich folgende Lessons Learned ableiten. Die Quizze werden in der genutzten Form (think-pair-share) sehr positiv bewertet. Zusätzlich sind sie wahrscheinlich mitverantwortlich dafür, dass die Studierenden sich relativ gut vorbereiten. Das Konzept ist deshalb möglicherweise auch in anderen ähnlichen Kontexten sinnvoll einsetzbar. Die Datenerhebungen waren hilfreich dabei, Konzepte und Fertigkeiten, mit denen viele Studierende Schwierigkeiten hatten, zu identifizieren, sodass Lehrende einschreiten und unterstützen konnten. Weiterhin ließen sich konkrete Ansatzpunkte für Verbesserungen daraus ableiten. Insofern können die Erhebungsmethoden im nächsten Durchlauf erneut verwendet werden.

# 6. Ausblick

Die dargelegten Erfahrungen und Erkenntnisse bilden aus unserer Sicht eine sehr gute Basis für folgende Iterationen. Aufgrund der Ergebnisse planen wir, das grundlegende Konzept (Literaturarbeit, Quizze, Übungsaufgaben) beizubehalten. Gleichzeitig konnten wir auf Basis der Daten eine Reihe von konkreten Verbesserungsmöglichkeiten erarbeiten, die in die Lehre einfließen werden. Ein eigenes Skript steht hierbei ein vorderster Stelle. Zusätzlich scheint es wichtig, die ersten Wochen weniger anspruchsvoll zu gestalten.

# Danksagung

# Literatur

Angaben zu den Autoren/Autorinnen

Ulrich von Zadow

Professor für Medieninformatik an der TH Nürnberg. Zuvor Studium der Informatik sowie langjährige Tätigkeit als Softwareentwickler und -entwicklungsleiter in der Exponatentwicklung für Museen und Ausstellungen. Anschließend Promotion im Bereich Mensch-Maschine Interaktion an der TU Dresden und Professur an der CODE University, Berlin. ORCID: https://orcid.org/0000-0001-7916-7433.

Natalie Kiesler

Natalie Kiesler ist Professorin für Lehren und Lernen an Hochschulen an der Fakultät Informatik der TH Nürnberg. Zuvor war sie leitende wissenschaftliche Mitarbeiterin am DIPF und Dozentin an der Goethe-Universität Frankfurt, wo sie 2022 in Informatik promovierte. Ihre Forschungsschwerpunkte umfassen Programmierkompetenz, Lernumgebungen und Feedback in der universitären Ausbildung. Aktuelle Projekte thematisieren generative KI in Bildungskontexten, Feedback, Open Science. ORCID: https://orcid.org/0000-0002-6843-2729.